\documentclass[aps,prl,twocolumn,groupedaddress]{revtex4}

\usepackage{graphicx}

\begin{document}

%\preprint{}

\title{Evidence for crossed Andreev reflection in superconductor-ferromagnet hybrid structures}


\author{D. Beckmann}
\email[e-mail address: ]{detlef.beckmann@int.fzk.de}
%\thanks{}
\author{H. B. Weber}
\affiliation{Forschungszentrum Karlsruhe, Institut f\"ur Nanotechnologie, P.O. Box 3640, D-76021 Karlsruhe, Germany}
\author{H. v. L\"ohneysen}
\affiliation{Forschungszentrum Karlsruhe, Institut f\"ur Festk\"orperphysik, P.O. Box 3640, D-76021 Karlsruhe, Germany,
and Physikalisches Institut, Universit\"at Karlsruhe, D-76128 Karlsruhe, Germany}


\date{\today}

\begin{abstract}
We have measured the non-local resistance of aluminum-iron spin-valve structures
fabricated by e-beam lithography and shadow evaporation. The sample geometry 
consists of an aluminum bar with two or more ferromagnetic wires forming
point contacts to the aluminum at varying distances from each other.
In the normal state of aluminum, we observe a spin-valve signal which allows
us to control the relative orientation of the magnetizations of the
ferromagnetic contacts. In the superconducting state, at low temperatures and
excitation voltages well below the gap, we observe a spin-dependent 
non-local resistance which decays on a smaller length scale than the 
normal-state spin-valve signal. The sign, magnitude and decay length of 
this signal is consistent with predictions made for 
crossed Andreev reflection (CAR).
\end{abstract}

%\keywords{}

\maketitle


Singlet superconductivity and ferromagnetism are competing long-range orders
which favor a different alignment of electron
spins, antiparallel and parallel, respectively. Therefore, 
they generally exclude each other in homogenous bulk materials.
In mesoscopic hybrid structures, the interplay of superconductivity and
ferromagnetism leads to rich novel physics.
Recent experimental studies on equilibrium properties of
superconductor-ferromagnet (SF) interfaces have shown that the local density of states (LDOS)
on the superconducting side is strongly affected
by the pair breaking effect of the ferromagnetic exchange field \cite{sillanpaa2001},
and on the ferromagnetic side an oscillatory behavior
of the LDOS due to the exchange splitting of the spin sub-bands has been
observed \cite{kontos2001}. Transport properties of SF point 
contacts \cite{soulen1998,upadhyay1998,perez2004} and FSF planar
junctions \cite{gu2002} show a suppression
of spin-polarized current injection into the superconductor, 
and in SFS Josephson junctions \cite{ryazanov2001,kontos2002,bauer2004},
a $\pi$-state with a spontaneous equilibrium Josephson current is observed. 
In this work, we report on spin-dependent
transport properties of the superconducting condensate on
length scales comparable to the coherence length. We have combined
ferromagnet-superconductor point contacts with a non-local spin-valve 
geometry, and find evidence for crossed Andreev 
reflection (CAR), i.e. the splitting of a Cooper pair into two
spatially separated leads. 

\begin{figure}
\includegraphics[width=\columnwidth]{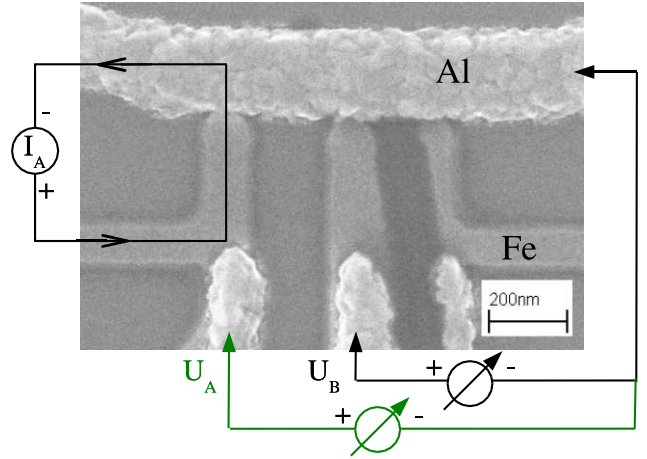}
\caption{\label{fig_sem}(Color online) SEM image of sample T2, and experimental scheme. 
Three vertical iron wires are connected by point contacts to 
a horizontal aluminum bar. The outer two iron wires have additional voltage
probes.
An example of the current injection ($I_\mathrm{A}$) and both local ($U_\mathrm{A}$) and 
non-local ($U_\mathrm{B}$)
voltage detection scheme for one pair of contacts is also shown.}
\end{figure}

Figure \ref{fig_sem} shows an SEM image of one of our 
samples, together with a schematic view of the experiment. 
The samples were fabricated by e-beam lithography and shadow evaporation
techniques. First, 20~nm of iron is evaporated onto an oxidized
silicon substrate to form a series of 
wires (vertical), with varying width (50-120 nm) and a small tip at the end. Their 
elongated shape creates a magnetic shape anisotropy 
which confines their magnetizations to be aligned along the wires, 
pointing either upwards or downwards, with different coercive fields
due to the width variation. In a second
evaporation step under a different angle, and without breaking the vacuum,
an aluminum bar of 80~nm height and 250~nm width (horizontal),
is created. It slightly touches the iron wires, forming metallic contacts of
about $20\times 50~\mathrm{nm}$,
much smaller than the dirty-limit coherence length $\xi\approx
200-300~\mathrm{nm}$ of the aluminum. This is crucial to
avoid a suppression of superconductivity by the proximity effect. 
A current $I_\mathrm{A}$ is injected through one of the contacts into the 
superconductor, and a voltage $U_\mathrm{B}$ is detected by a second contact 
outside the current path, defining the non-local resistance
$R=U_\mathrm{B}/I_\mathrm{A}$.
Samples of two different layouts were investigated. Layout T (shown)
has three iron wires and additional voltage probes for detection of the 
local resistance $U_\mathrm{A}/I_\mathrm{A}$. 
Layout S has six iron wires over a larger
range of distances, and no additional voltage probes. 
For each sample, different combinations of contacts have been used as
injector-detector pair in order to study the dependence of $R$ on the
distance $d$ between the two contacts. 
Here, we show data from one sample of each layout, T2 and S5.
The samples were mounted into a shielded box thermally anchored to the mixing
chamber of a dilution refrigerator. The measurement lines were fed through
a series of filters to eliminate RF and microwave radiation from the 
shielded box. Resistance was measured 
with a low-frequency AC resistance bridge, with an RMS excitation
amplitude $I_\mathrm{exc}$.
For differential resistance measurements, an additional DC bias current 
was applied.

\begin{figure}
\includegraphics[width=\columnwidth]{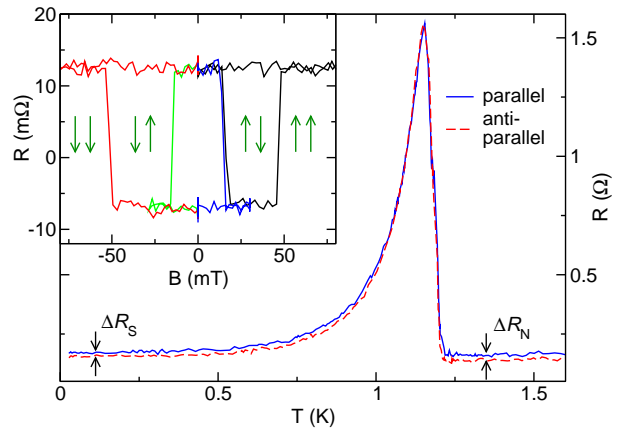}
\caption{\label{fig_t}(Color online)
Non-local resistance $R$ as a function of the temperature $T$ 
at zero external magnetic field for two contacts of sample T2 
at a distance $d=310~\mathrm{nm}$. The solid and dashed line corresponds to
parallel and antiparallel alignment of the injector and detector
magnetizations, respectively.\\
Inset: Non-local resistance $R$ at $T=1.8~\mathrm{K}$ as a function of a magnetic field $B$
applied parallel to the iron wires,
for two contacts of sample S5 at a distance $d=210~\mathrm{nm}$.
The resistance jumps correspond to magnetization reversals of the 
ferromagnetic
contacts. The arrows indicate the four different magnetization states.}
\end{figure}

For each pair of contacts, we first characterize the spin-valve behavior in the
normal state. The inset of Fig. \ref{fig_t} shows the non-local resistance for two
contacts as a function of the magnetic field $B$ applied along the
direction of the ferromagnetic wires at a temperature $T=1.8~\mathrm{K}$,  
with an excitation amplitude $I_\mathrm{exc}=1~\mathrm{\mu A}$. 
For each direction of the magnetic field, the
data show two sharp stepwise changes of the resistance which
correspond to the reversal of the magnetizations of the two contacts
at their individual coercive fields, yielding each two states
for parallel and antiparallel magnetization alignment.
These states remain stable at zero external magnetic field. 
The resistance difference $\Delta R_\mathrm{N}$ between parallel and
antiparallel alignment in the normal state is 
governed by the injection of non-equilibrium magnetization into aluminum,
and the associated  splitting of the chemical potentials for spin 
up and down electrons \cite{johnson1985}.
A quantitative analysis of $\Delta R_\mathrm{N}$ will be given below.

For each of the four magnetization states of a given contact pair,
the sample is cooled through the superconducting transition
at zero external magnetic field, and subsequently warmed up again.
Here a very small excitation amplitude $I_\mathrm{exc}=50~\mathrm{nA}$ is chosen
to avoid both quasiparticle injection above the superconducting gap and
self-heating. At this small excitation (and correspondingly small
detector voltage amplitude $U_\mathrm{B}\approx 0.5~\mathrm{nV}$), 
the limited absolute accuracy of the resistance bridge
yields an artificial offset of $\approx 150~\mathrm{m\Omega}$ to $R$.
The observed data depend only on the relative magnetization alignment of the two
Fe wires, and are 
therefore averaged over parallel and antiparallel configurations. 
The results for one pair of contacts is shown in Fig. \ref{fig_t}.
Above the superconducting transition temperature, the signal is constant,
with the well resolved difference $\Delta R_\mathrm{N}$ between
parallel and antiparallel alignment.
At the superconducting transition at $T_\mathrm{c}\approx 1.15~\mathrm{K}$, 
a large peak is observed, which is essentially independent of 
the magnetization alignment.
This peak is due to charge imbalance created by the injection of quasiparticles 
above the superconducting gap \cite{clarke1972,tinkham1972}, which is still small 
close to $T_\mathrm{c}$. 
As the gap opens further towards lower temperatures,
the charge-imbalance peak subsides, and finally, 
the resistance becomes constant again at $T<250~\mathrm{mK}$.
In this regime, the signal for parallel alignment is still
somewhat larger than for the antiparallel case. 
The difference 
$\Delta R_\mathrm{S}$ between the non-local resistances for
parallel and antiparallel alignment in the superconducting state
is the central result of this work, and we will now further clarify
its physical origin.

\begin{figure}
\includegraphics[width=\columnwidth]{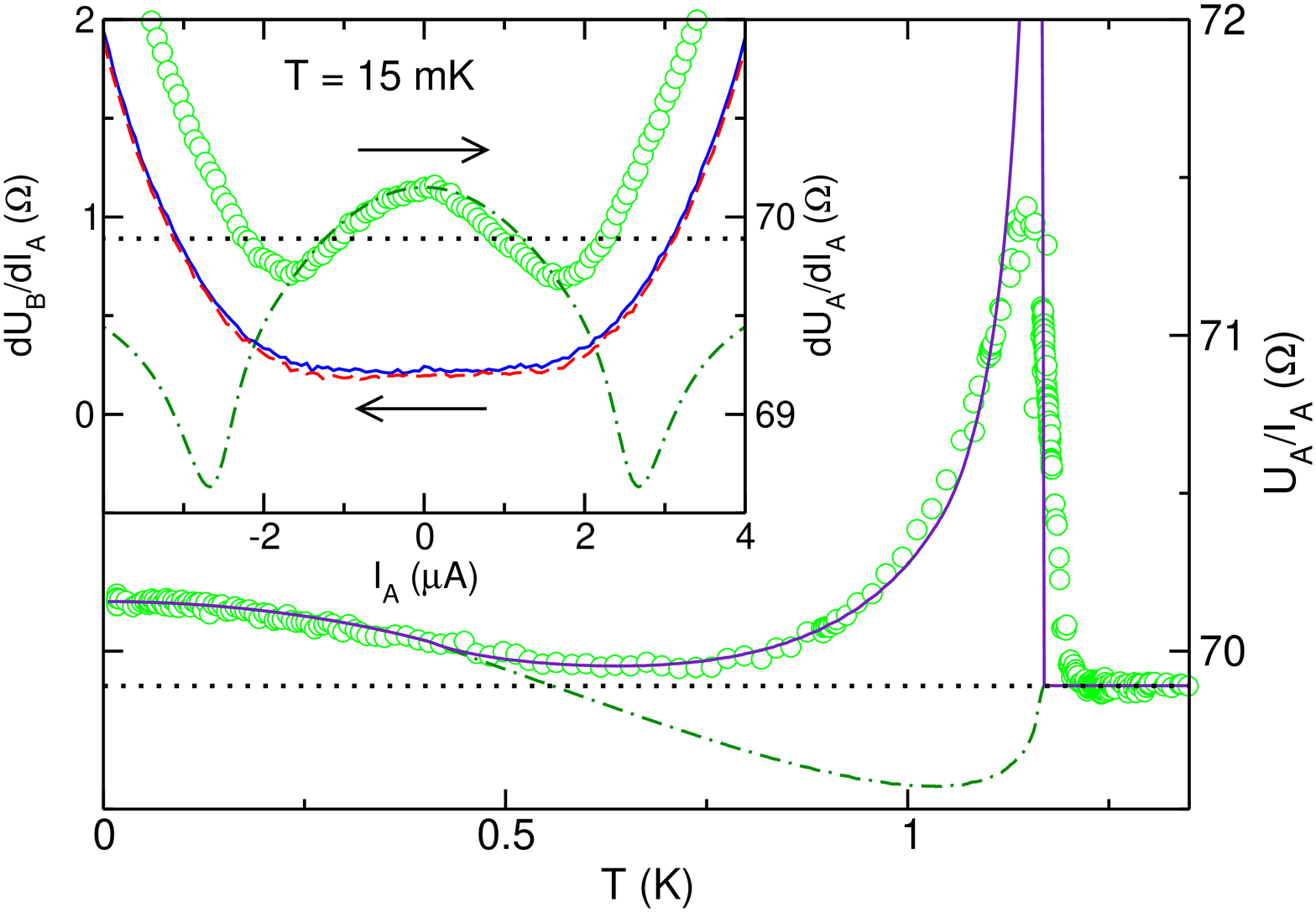}
\caption{\label{fig_iv}(Color online)
Symbols: Local resistance $U_\mathrm{A}/I_\mathrm{A}$ vs. temperature $T$ for
a contact of sample T2. Solid line: fit to
$R_0+R_\mathrm{BTK}(T)+R_\mathrm{ci}(T)$ (see text). Dash-dotted line:
the contribution $R_0+R_\mathrm{BTK}(T)$.\\
Inset: Differential local resistance $dU_\mathrm{A}/dI_\mathrm{A}$ vs. 
$I_\mathrm{A}$ for the same contact (symbols), and differential non-local
resistance $dU_\mathrm{B}/dI_\mathrm{A}$ for a detector at distance $d=210$~nm
(solid line: parallel, dashed line: antiparallel alignment). 
The dash-dotted line is the fit 
$R_0+R_\mathrm{BTK}(U_\mathrm{A})$ 
(see text), the dotted line indicates $R_0+R_\mathrm{pc}$ in both frames.
}
\end{figure}

A possible source of the spin-dependent signal in the superconducting
state are quasiparticles propagating above the superconducting 
gap. Such a signal has been observed in the charge-imbalance region 
near $T_\mathrm{c}$ using current
injection into a niobium film through a tunnel junction \cite{johnson1994}.
It was found to be an order of magnitude smaller than the normal-state 
spin-valve signal at a temperature 0.3\% below $T_\mathrm{c}$, 
and became unobservably small about 1.5\% below $T_\mathrm{c}$.
We do not observe such a signal close to $T_\mathrm{c}$, presumably because
it is buried in the steep spin-independent increase caused by charge imbalance. 

To check further whether $\Delta R_\mathrm{S}$ observed at low temperature
is due to quasiparticles above the gap, or due to sub-gap transport, we have
employed the additional voltage probes of sample T2 to 
characterize the point contacts. 
The local resistance $U_\mathrm{A}/I_\mathrm{A}$ of one contact is shown
in Fig. \ref{fig_iv} as a function of temperature.
The data can be described by $R(T)=R_0+R_\mathrm{BTK}(T)+R_\mathrm{ci}(T)$,
where $R_0=63.7~\mathrm\Omega$ is the resistance of the ferromagnetic strip 
leading to the point-contact, $R_\mathrm{BTK}(T)$ is the temperature dependence 
of the point-contact resistance according to
a spin-polarized extension of the BTK theory \cite{blonder1982,strijkers2001},
and $R_\mathrm{ci}(T)$ is a contact resistance due to charge imbalance 
\cite{chi1979,chi1980,hsiang1980,blonder1982},
which is essentially the same as observed in the non-local resistance.
Parameters for the BTK model are $R_\mathrm{pc}=6.2~\mathrm{\Omega}$, $P=0.45$,
$Z=0.22$ and $\Delta=177~\mathrm{\mu eV}$ for the normal state point-contact 
resistance, the interfacial spin polarization, the interface 
transparency, and the gap, respectively. The fit to our model is excellent
except for a slight broadening around $T_\mathrm{c}$.

The low bias part ($|I_\mathrm{A}| < 1.7~\mathrm{\mu A}$) of the differential local 
resistance $dU_\mathrm{A}/dI_\mathrm{A}$ can be well described by 
$R_0+R_\mathrm{BTK}(U_\mathrm{A})$ using 
$U_\mathrm{A}=I_\mathrm{A}(R_0+R_\mathrm{pc})$
(i.e. assuming insignificant inelastic scattering in the Fe strip) 
and the same parameters as for the 
temperature dependence, whereas the differential non-local resistance 
$dU_\mathrm{B}/dI_\mathrm{A}$ remains
constant in that bias region (inset of Fig. \ref{fig_iv}).
At higher bias $|I_\mathrm{A}| > 1.7~\mathrm{\mu A}$,
both local and non-local differential resistance show a steep increase due to
charge imbalance, i.e. quasiparticle injection above the gap. 
The apparent reduction of the gap compared to the BTK prediction can 
be attributed to self-heating at higher bias.
We conclude that the constant $\Delta R_\mathrm{S}$ observed in the
low-bias region is due to sub-gap transport. 

Sub-gap transport between normal metals and superconductors is mediated
by Andreev reflection \cite{andreev1964}:
An electron from the normal metal enters the superconductor,
and a hole of opposite spin is retro-reflected, creating a spin-singlet
Cooper pair in the superconductor. 
In a multi-terminal structure,
it has been predicted that the incident electron and the retro-reflected hole 
may be transmitted through 
different contacts (crossed Andreev reflection, or CAR),
 as long as the distance
between the two contacts does not exceed the superconducting coherence
length \cite{byers1995,deutscher2000}, 
yielding a negative non-local resistance. There is a second non-local process, 
electron cotunneling (EC), where an electron enters the superconductor 
through one contact, 
and an electron of the same spin leaves through the second contact,
yielding positive non-local resistance. 
If the two contacts are spin polarized, EC and CAR are
favorable for parallel and antiparallel magnetization alignment, respectively.
We propose the superposition of
CAR and EC processes as the origin of the observed non-local resistance
difference in the superconducting state.

\begin{figure}
\includegraphics[width=\columnwidth]{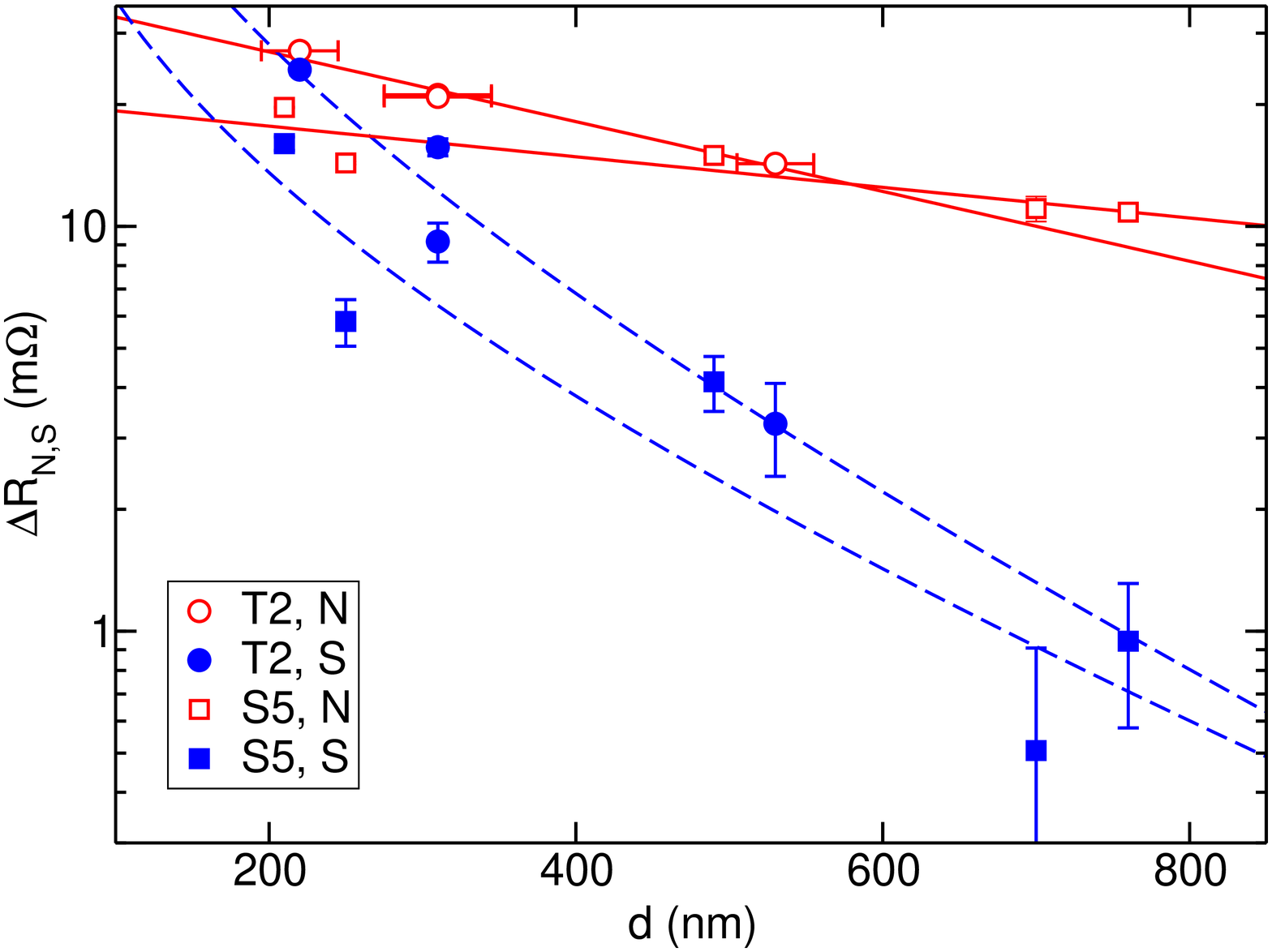}
\caption{\label{fig_d}(Color online)
Difference $\Delta R_\mathrm{N,S}$ between parallel and antiparallel alignment vs. distance
$d$ between the contacts, for
two samples in both the normal (open symbols) and superconducting (closed
symbols) state.
The solid and dashed lines are fits to (\ref{equ_sv}) and (\ref{equ_ca}), 
respectively, as described in the text.}
\end{figure}

To obtain a quantitative description of our results, and to test our
hypothesis, we have 
extracted $\Delta R_\mathrm{N}$ and 
$\Delta R_\mathrm{S}$ from the raw data and 
investigated their dependence on the distance $d$ between the 
injector and detector contact. These data are shown in Fig. \ref{fig_d}. 
Both signals are of similar magnitude at short distance, and decay
at increasing $d$. The decay in the superconducting state takes place
on a much shorter length scale than in the normal state,
indicating the different transport mechanism.

For the normal state data, we use a semiclassical model
of spin injection and diffusion \cite{takahashi2003}.
The key parameters are the bulk spin polarization
$p=(\sigma^{\mathrm{F}\uparrow}-\sigma^{\mathrm{F}\downarrow})/(\sigma^{\mathrm{F}\uparrow}+\sigma^{\mathrm{F}\downarrow})$
of the ferromagnet, where $\sigma^{\mathrm{F}\uparrow}$ and $\sigma^{\mathrm{F}\downarrow}$
are the spin-dependent conductivities of the ferromagnet, and the quantities 
$\mathcal{R}^\mathrm{N}=\rho^\mathrm{N}\lambda_\mathrm{s}^\mathrm{N}/A^\mathrm{N}$
and 
$\mathcal{R}^\mathrm{F}=\rho^\mathrm{F}\lambda_\mathrm{s}^\mathrm{F}/A^\mathrm{J}$,
where $\rho^\mathrm{(N,F)}$ and $\lambda_\mathrm{s}^\mathrm{(N,F)}$ 
are the (spin-averaged) resistivity and spin-diffusion
length in the normal metal and ferromagnet, respectively, 
$A^\mathrm{N}$ is the cross section of the normal metal and
$A^\mathrm{J}$ is the contact area. For our samples, we estimate
$(\mathcal{R}^\mathrm{F}\approx 3\Omega) > (\mathcal{R}^\mathrm{N}\approx 1\Omega)$.
In that limit, and neglecting contact resistances and
inhomogenous current distributions, $\Delta R_\mathrm{N}$ is given by
\begin{equation}
\label{equ_sv}
\Delta R_\mathrm{N} \approx p^2
\frac{\rho^\mathrm{N}\lambda_\mathrm{s}^\mathrm{N}}{A^\mathrm{N}} \exp( - d /
\lambda_\mathrm{s}^\mathrm{N} ).
\end{equation}

%\begin{table}%[H] add [H] placement to break table across pages
%\caption{\label{tab_exp}
%Measured normal state resistivity $\rho^\mathrm{N}$ of the aluminum, 
%and parameters obtained from the fits shown in Fig. \ref{fig_d},
%as described in the text.
%}
%\begin{ruledtabular}
%\begin{tabular}{lrrl}
%sample                          & T2   & S5    &             \\ \hline
%$\rho^\mathrm{N}$               & 2.85 & 1.41  & $\mu\Omega$cm \\ \hline
%$A$		                & 40.1 m$\Omega$     & 21.0 m$\Omega$     \\
%$R^\mathrm{N}                  & 0.82               & 0.84               \\
%$p$                             & 22   & 16    & \%              \\
%$\lambda_\mathrm{s}^\mathrm{N}$ & 505  & 1150  & nm            \\ \hline
%$\tau_\mathrm{s}^\mathrm{N}$    & 47   & 96    & ps            \\
%$P$                             & 51   & 47    & \%              \\
%$\xi$                           & 275  & 345   & nm             \\
%\end{tabular}
%\end{ruledtabular}
%\end{table}

A joint description of both local and non-local sub-gap conductance in the
superconducting state is given by \cite{falci2001}
\begin{equation}
\label{equ_conductance}
\left(\begin{array}{c} I_\mathrm{A} \\ I_\mathrm{B}\end{array}\right) 
=
\left(\begin{array}{cc} G_\mathrm{A} & G_\mathrm{CAR}-G_\mathrm{EC} \\ 
G_\mathrm{CAR}-G_\mathrm{EC} & G_\mathrm{B}
\end{array}\right)
\left(\begin{array}{c} U_\mathrm{A} \\ U_\mathrm{B}\end{array}\right),
\end{equation}
where $I_\mathrm{A,B}$, $U_\mathrm{A,B}$ and $G_\mathrm{A,B}$ are the 
current, voltage and local conductance of the injector (A) and
detector (B) contact, $G_\mathrm{CAR}$ and $G_\mathrm{EC}$ are the
non-local conductances due to CAR and EC processes, and using
$G_\mathrm{A},\,G_\mathrm{B}\gg G_\mathrm{CAR},\,G_\mathrm{EC}$ throughout. 
Inverting (\ref{equ_conductance}) 
for the case $I_\mathrm{B}=0$ (voltage detection at B) we find
the non-local resistance
\begin{equation}
\label{equ_r}
R_\mathrm{S}=U_\mathrm{B}/I_\mathrm{A}\approx -(G_\mathrm{CAR}-G_\mathrm{EC})/(G_\mathrm{A}G_\mathrm{B}).
\end{equation}
Quantitative predictions for a diffusive superconductor yield, to
lowest order in transmission coefficients \cite{feinberg2003}
\begin{equation}
\label{equ_feinberg}
\left(\begin{array}{c} G_\mathrm{CAR} \\ G_\mathrm{EC}\end{array}\right)
\approx\frac{\pi}{4}
\left(\begin{array}{c} 
G_\mathrm{A}^{\uparrow}G_\mathrm{B}^{\downarrow}+G_\mathrm{A}^{\downarrow}G_\mathrm{B}^{\uparrow} \\ 
G_\mathrm{A}^{\uparrow}G_\mathrm{B}^{\uparrow}+G_\mathrm{A}^{\downarrow}G_\mathrm{B}^{\downarrow}
\end{array}\right)
\frac{\exp(-d/\xi)}{Ne^2Dd}, 
\end{equation}
where $G_\mathrm{A,B}^{\uparrow,\downarrow}$ are the spin-dependent
contact conductances, and $N$, $D$ and $\xi$ are the density of states at the 
Fermi level, the diffusion
coefficient and the dirty limit coherence length of aluminum, respectively.
Using Einstein's relation $Ne^2D=1/\rho^\mathrm{N}$, we find
$R_\mathrm{S} \approx \pm
P_\mathrm{A}P_\mathrm{B}\pi\rho^\mathrm{N}\exp(-d/\xi)/4d$,
where $+$ and $-$ are for parallel and antiparallel configuration,
and $P_\mathrm{A,B}=(G_\mathrm{A,B}^\uparrow-G_\mathrm{A,B}^\downarrow)/G_\mathrm{A,B}$
are interfacial spin polarizations, and with
$P=P_\mathrm{A}=P_\mathrm{B}$
\begin{equation}
\label{equ_ca}
\Delta R_\mathrm{S}\approx
P^2\frac{\pi}{2}\rho^\mathrm{N}\frac{\exp(-d/\xi)}{d}.
\end{equation}
Recent predictions for high transparency interfaces indicate the same 
distance dependence and an amplitude within a factor of three of the
low-transparency case \cite{melin2004}.

Fits of (\ref{equ_sv}) and (\ref{equ_ca}) to our data in the normal and 
superconducting state, respectively, are shown in Fig. \ref{fig_d}.
Within the experimental scatter, the data are described well by the
fits.
We find $\lambda_\mathrm{S}^\mathrm{N}=500-1000$~nm, consistent with
values for aluminum obtained by spin-valve experiments \cite{jedema2003},
and a bulk spin polarization $p\approx 20$~\%.
The interfacial spin polarization $P\approx 50$~\% 
obtained from the fits in the superconducting
state  is in excellent agreement with $P\approx 45$~\%
obtained from point contact spectroscopy on iron \cite{soulen1998,strijkers2001},
and $\xi\approx 300~\mathrm{nm}$ is consistent with
the dirty limit predictions calculated from the resistivity.
Thus, our data in the superconducting state are in good quantitative agreement 
with the predicted behavior.

To conclude, we have investigated experimentally non-local
sub-gap transport in superconductors-ferromagnet hybrid structures, and 
shown evidence for the superposition of crossed Andreev reflection (CAR)
and electron cotunneling. 
CAR has been discussed as
a sensitive probe of superconducting order parameters \cite{byers1995}. 
In our case of a dirty (hence isotropic) BCS superconductor, 
the order parameter enters only through an exponential decay with $\xi$. 
With sufficient spatial resolution an investigation of
unconventional superconductors appears feasible. More recent proposals
in the context of quantum information processing
(see \cite{recher2003} and references therein)
have pointed out that CAR, if viewed as the decay of a Cooper
pair into two electrons in different normal metal leads, 
creates a spatially separated entangled electron pair. This entanglement 
can be probed by measuring noise cross-correlations in our structures.

\begin{acknowledgments}
Useful discussions with B. J. van Wees, C. Strunk and M. Aprili are gratefully
acknowledged. This work was supported by the Deutsche Forschungsgemeinschaft
within the framework of the Center for Functional Nanostructures.
\end{acknowledgments}

\bibliography{care.bib,../../lit.bib}

\end{document}